\documentclass[11pt,draftcls,onecolumn] {IEEEtran}

\usepackage[cmex10]{amsmath}
\usepackage{array}

\usepackage{amssymb}
\usepackage{amsmath}
\usepackage{amsthm}
\usepackage[lined,boxed,commentsnumbered, ruled]{algorithm2e}
\usepackage{xcolor} 

\usepackage{graphicx}
\usepackage{epstopdf}

\ifCLASSOPTIONcompsoc
  \usepackage[caption=false,font=normalsize,labelfont=sf,textfont=sf]{subfig}
\else
  \usepackage[caption=false,font=footnotesize]{subfig}
\fi

\begin{document}

\title{Performance of ML Range Estimator in Radio Interferometric Positioning Systems}
%
%

\author{Yue~Zhang,
        Wangdong~Qi*,~\IEEEmembership{Member,~IEEE},
        Guangxia~Li,
        Su~Zhang
\thanks{This work was supported by the National Natural Science Foundation of China (Grants 61273047 and 61301159) and the Natural Science Foundation of Jiangsu Province, China (BK20130068).}
\thanks{The authors are with the PLA University of Science and Technology, Nanjing, Jiangsu 210007, China (e-mail: zhyemf@gmail.com; wangdongqi@ gmail.com; 13905177686@139.com; franklinzhang1985@gmail.com).}
}

%
%

\markboth{submitted to ieee signal processing letters,~Vol.~X, No.~X, X~2014}%
{yue zhang \MakeLowercase{\textit{et al.}}: Performance of ML Range Estimator in Radio Interferometric Positioning Systems}
%

\maketitle


\begin{abstract}
The radio interferometric positioning system (RIPS) is a novel positioning solution used in wireless sensor networks. This letter explores the ranging accuracy of RIPS in two configurations. In the linear step-frequency (LSF) configuration, we derive the mean square error (MSE) of the maximum likelihood (ML) estimator. In the random step-frequency (RSF) configuration, we introduce average MSE to characterize the performance of the ML estimator. The simulation results fit well with theoretical analysis. It is revealed that RSF is superior to LSF in that the former is more robust in a jamming environment with similar ranging accuracy.
\end{abstract}


%
\IEEEpeerreviewmaketitle

\section{Introduction}
\IEEEPARstart{T}{he} radio interferometric positioning system (RIPS), a node localization system used in wireless sensor networks, has received significant attention in recent years \cite{maroti2005radio}, \cite{wang2013design}, \cite{li2013distance}, \cite{kusy2007intrack}, \cite{liu2011ground}. The novel ranging scheme introduced in RIPS is key to its success in providing low-cost and accurate localization solutions. However, our knowledge of the ranging performance of RIPS is rather limited because of the lack of systematic investigation. This letter examines its performance in two measurement configurations including linear step-frequency (LSF) and random step-frequency (RSF). LSF is applied in mobile node tracking and landslide early warning systems \cite{kusy2007intrack}, \cite{liu2011ground} whereas RSF can be essential for military applications with inherent anti-jamming capabilities \cite{weili}.

The topic of interest here is the ranging accuracy of RIPS both in high signal-to-noise ratio (SNR) and moderate--low SNR since RIPS may be deployed in a variety of environments. Because the performance predictions provided by lower bounds such as the Cramer-Rao bound (CRB) and the Ziv-Zakai bound (ZZB) are too optimistic when the SNR is below a certain threshold \cite{athley2005threshold}, \cite{richmond2006mean}, we choose to characterize the ranging performance of RIPS by using the mean square error (MSE) of the maximum likelihood (ML) estimator.

To obtain the MSE of the ML estimator in the entire SNR region,
we employ the method of interval error (MIE) \cite{athley2005threshold}, \cite{richmond2006mean}. In the LSF configuration, the measurement frequencies are fixed so we can obtain the MSE with MIE directly. In the RSF configuration, because the MSE is a random variable with respect to hopping frequencies in measurement, we use the average MSE (AMSE) \cite{MIMOradar07}  to characterize the performance of the ML estimator. We introduce the average ambiguity function (AAF) \cite{kaveh1976average}  to facilitate the derivation of AMSE to avoid the tedious process of averaging MSEs under different measurement frequencies.

The theoretical results are verified by simulations. The ranging accuracy of RSF is shown to be very similar to that of LSF. Therefore, RSF is superior to LSF for military applications because it is more robust in a jamming environment.

\section{System Model}
The basic unit of a ranging process in RIPS involves two nodes, A and B, simultaneously emitting a pair of sine waves at two close frequencies with a difference of $\delta$, whereas other two nodes, C and D, measure the phase of the beat signal of the two sine waves. The overall ranging process consists of multiple such units at a series of frequency pairs. According to \cite{maroti2005radio}, the phase offset $\varphi_i$ between C and D of the $i$th beat signal is related to the so-called \emph{qrange} $d_0=d_{AD}-d_{AC}+d_{BC}-d_{BD}$ ($d_{XY}$  is the distance between node $X$ and $Y$) as
\begin{equation}\label{phii}
  \varphi_i = \left(2\pi\frac{f_i}{c}d_0 + \theta + n_i \right)\;\mod 2\pi,
\end{equation}
where $f_i$ is the average of the $i$th pair of frequencies ($i=1,\cdots ,M$), $c$ is the speed of signal propagation, $n_i$ is independent and identically distributed (i.i.d) Gaussian white noise with variance $\sigma^2$, and $\theta=\frac{2\pi\delta}{c}(d_{AD}-d_{AC}-d_{BC}+d_{BD})$ is a constant related to $d_0$. We define the SNR as $1/\sigma^2$.

Essentially, the ranging process in RIPS distills to a parameter estimation problem where the qrange $d_0$, a linear combination of distances between the four nodes, is determined according to the observation of the phase differences $\mathbf{\Phi}=\{\varphi_1,\cdots,\varphi_M \}$ of the beat signals on measurement frequencies $\mathbf{f}=\{f_1,\cdots ,f_M \}$.

It should be noted here that qranges can be used with ease in the localization process in similar ways like distances. Due to space limitations, we refer the readers to \cite{kusy2007intrack} for further information.

The observation equation (\ref{phii}) has been simplified in previous study by neglecting the term $\theta$ \cite{maroti2005radio}. In this letter, we retain the general form of (\ref{phii}) by treating $\theta$  as an unknown parameter to accommodate additional scenarios.

We assume that all measurement frequencies employed in the ranging process are multiples of the system's minimum frequency interval $f_{min}$
\begin{equation}\label{fi}
  f_i = (k_0+ k_i)f_{min},
\end{equation}
where $k_0 f_{min}$ is the initial frequency. Assuming that measurement frequencies are chosen from the available bandwidth $B$ in RIPS, it is clear that the total number of measurement frequencies is $N=B/f_{min}+1$.

Obviously, the ranging process in RIPS is defined by the configuration of measurement frequencies $\mathbf{f}$. In the LSF configuration, $f_i$ proceeds in a constant step, i.e., $k_i=(i-1)\frac{N-1}{M-1}$. We assume that $N-1$ can be divided by $M-1$ for convenience. In the RSF configuration, the $M$ measurement frequencies are chosen randomly from all available frequencies so that the positive integers $k_i$ are random variables distributed uniformly in $[0,N-1]$.

\section{Performance Analysis}
We first present the ML estimator of qrange in RIPS. In a fairly large SNR region, the observation equation (\ref{phii}) can be converted into an equivalent form \cite{tretter1985estimating}, \cite{awoseyila2008improved}:
\begin{equation}\label{exp_phii}
  \exp(j\varphi_i) = \exp\left(j(2\pi\frac{f_i}{c}d_0 + \theta) \right) + z_i,
\end{equation}
where $z_i$ is i.i.d complex Gaussian white noise with variance $2\sigma^2$ corresponding to the additive phase noise $n_i$ with variance $\sigma^2$ in (\ref{phii}). The estimation of qrange $d_0$ is equivalent to single-tone frequency estimation if we regard $d_0$ as the frequency of a single tone and ${f_i}/{c}$ as the discrete sample time. Then the joint distribution function of $\exp(j\mathbf{\Phi})$ at $\mathbf f$ with the unknown parameter vector $\mathbf{A}=[d_0,\theta]^T$ is \cite{rife1974}
\begin{equation}\label{f_ml}
\begin{array}{cc}
   f(\exp(j\mathbf{\Phi});\mathbf{A}) = \left(\tfrac{1}{2\pi \sigma^2}\right)^M &\exp \left[-\tfrac{1}{2\sigma^2}\left(\sum_{i=1}^{M}(a_i-\mu_i)^2\right.\right. \\ 
    &+ \left.\left.\sum_{i=1}^{M}(b_i-v_i)^2\right) \right],
\end{array}
\end{equation}
where $a_i=\mathbf{Re}(\exp j\varphi_i)$, $b_i=\mathbf{Im}(\exp j\varphi_i)$, $\mu_i = \cos(2\pi\frac{f_i}{c}d_0 + \theta)$, and $v_i = \sin(2\pi\frac{f_i}{c}d_0 + \theta)$. As a result, the ML estimate $\hat d_0$ is obtained by maximizing the objective searching function (OSF)
\begin{equation}\label{v_daijiahanshu}
  V(d) = \left| \sum_{i=1}^{M}\left\{ \exp(j\varphi_i)\exp(-j2\pi\tfrac{f_i}{c}d) \right\}\right|,
\end{equation}
where $d\in \left[ d_0-\frac{d_{max}}{2}, d_0+\frac{d_{max}}{2}\right]$, and $d_{max}$ is the range of interest within the unambiguous distance of RIPS.

According to MIE \cite{athley2005threshold}, \cite{richmond2006mean}, we represent the MSE of the ML qrange estimator in configuration $\mathbf{f}$ as a weighted sum of the local error term and the global error term (outlier)
\begin{equation}\label{MSE_f}
  MSE(d_0|_\mathbf{f}) = P_{o} \cdot E[(\hat{d_0}-d_0)^2|outlier] +　(1-P_{o}) \cdot CRB(d_0|\mathbf{f}),
\end{equation}
where the weights are given by the outlier probability $P_o$ and the local error is approximated by the CRB term $CRB(d_0|\mathbf{f})$.


Next, we handle LSF and RSF in sections \ref{theLSFconfiguration} and \ref{theRSFconfiguration}.

\subsection{The LSF Configuration}\label{theLSFconfiguration}
For the LSF configuration, (\ref{MSE_f}) can be simplified further if we introduce the concept of the ambiguity function (AF), which is the OSF when the data in (\ref{v_daijiahanshu}) is noise free. According to \cite{richmond2006mean}, an outlier is an event that occurs when the ML parameter estimate is outside the mainlobe of the AF. The AF can be discretized at the sidelobe peaks $d_n(n=0,1,\cdots,N_p)$, where $d_n$s are positions and $N_p$ is the number of sidelobe peaks of the AF. Under this discretization, $P_o$ and the first term in (\ref{MSE_f}) can be simplified as \cite{athley2005threshold}
\begin{equation}\label{Po1}
  P_{o} \approx \sum_{n=1}^{N_p} p_n
\end{equation}
and
\begin{equation}\label{out1}
  P_{o} \cdot E[(\hat{d_0}-d_0)^2|outlier] \approx \sum_{n=1}^{N_p}  p_n(d_n - d_0)^2,
\end{equation}
where $p_n = \Pr[ V(d_0)< V(d_n)]$ is the probability that the  sidelobe peak of OSF at $d_n$ is higher than the mainlobe.

Combining (\ref{MSE_f}), (\ref{Po1}) and (\ref{out1}), the MSE of the ML qrange estimation in LSF can be approximated as
\begin{equation}\label{MSE_final}
  MSE_{LSF} \approx \sum_{n=1}^{N_p}p_n(d_n-d_0)^2 + \left(1 - \sum_{n=1}^{N_p}p_n\right)CRB_{LSF},
\end{equation}
where $CRB_{LSF}$ represents $CRB(d_0|\mathbf{f})$ in the LSF configuration. We now address the determination of terms in (\ref{MSE_final}).

\subsubsection{CRB}
The elements of the Fisher information matrix $\mathbf{J}$ corresponding to (\ref{f_ml}) can be written as
\begin{equation}\label{Jii}
  \mathbf{J}_{ij} = -E\left[\frac{\partial ^2 \ln f(\exp(j\mathbf{\Phi});\mathbf{A})}{\partial A_i \partial A_j}\right].
\end{equation}
Inverting $\mathbf{J}$  yields CRB for the ML estimator of $d_0$ such that
\begin{equation}
    \textstyle CRB(d_0|\mathbf{f}) = \mathbf{J}^{-1}_{11}=\frac{M c^2 \sigma^2}{4\pi^2} \Big/ \left[M\sum_{i=1}^{M}{f_i}^2 - \left(\sum_{i=1}^{M}f_i\right)^2\right].
\end{equation}
Replacing $f_i$ with the right hand side of (\ref{fi}), we get
\begin{equation}\label{CRB_f}
  CRB(d_0|\mathbf{f}) = \frac{M c^2 \sigma^2}{4\pi^2 f_{min}^2} \frac{1}{\mathbf{K}^{T}\mathbf{W K}},
\end{equation}
where $\mathbf{K}=[k_1,\dots,k_{M}]^T$, $\mathbf{W}$ is an $M\times M$ symmetric matrix with main diagonal elements $ M-1$ and others $-1$.

Considering that $k_i$ increases stepwise by $\frac{N-1}{M-1}$ in the LSF configuration, we get from (\ref{CRB_f})
\begin{equation}\label{CRB_final}
  CRB_{LSF} = \frac{3 c^2 \sigma^2 (M-1)}{\pi^2 B^2 M(M+1)}.
\end{equation}

\subsubsection{Outlier Related Terms}
From the ambiguity function
\begin{equation}
  \begin{array}{cc}
     G(d)=\left |\sum_{i=1}^{M} \exp\left( j2\pi \frac{f_i}{c}(d_0 - d)\right)\right |\\
         =\left| \frac{\sin\left(\pi(d_0-d)B\frac{M}{(M-1)c}\right)}{\sin\left(\pi(d_0-d)\frac{1}{(M-1)c}\right)} \right|,
  \end{array}
\end{equation}
we get $N_p=M-2$ and $d_n=d_0 +(-1)^n \tfrac{(M-1)c}{MB}(\lceil N/2\rceil+0.5)$.

Let
\begin{equation}
  \begin{array}{ll}
  y_0 = \sum_{i=1}^{M}\exp(jn_i)\\
  y_n = \sum_{i=1}^{M}\exp\left(j2\pi\tfrac{f_i}{c}(d_0-d_n) + jn_i\right);
  \end{array}
\end{equation}
we have $V(d_0)=|y_0|$, $V(d_n)=|y_n|$, and
\begin{equation}
  p_n=\Pr(|y_0|-|y_n|<0) = \Pr(|y_0|^2-|y_n|^2<0).
\end{equation}

It is observed that $y_0$, as well as $y_n$, is the sum of $M$ i.i.d random variables. In view of central-limit theorem, both $y_0$ and $y_n$ are approximately Gaussian distributed if $M$ is sufficiently large. In addition, because $y_0$ and $y_n$ are correlated, we can get the expression of $p_n$ resorting to appendix B in \cite{proakis} by means of the first- and second-order moments of $y_0$ and $y_n$.

It should be noted that if $x$ is normally distributed, we have $E[\exp(jx)]=e^{-\sigma^2/2}$ according to the definition of the character function \cite{papoulis}. Then, routine computation produces the first- and second-order moments of $y_0$ and $y_n$:
\begin{equation}\label{moments}
 \begin{array}{ll}
    E[y_0] = M e^{-\sigma^2/2}\\
    E[y_n] = M r_n e^{-\sigma^2/2}\\
    var[y_0] =var[y_n] = M (1-e^{-\sigma^2})\\
    cov[y_0,y_n] = M ({r_n})^* (1-e^{-\sigma^2}),
 \end{array}
\end{equation}
where $r_n =\frac{1}{M} \sum_{i=1}^{M}\exp \left( j2\pi \tfrac{f_i}{c}(d_0-d_n)\right)$ is the relative sidelobe level of the $n$th sidelobe of the ambiguity function \cite{athley2005threshold}, and the superscript $(\cdot)^*$ means conjugation.

Substituting (\ref{moments}) into B-21 of \cite{proakis}, we have
\begin{equation}\label{pn}
  p_n = Q_1(a,b) - \frac{1}{2} I_0(ab) \exp\left[-\tfrac{1}{2}(a^2+b^2)\right].
\end{equation}
Here, $Q_1(\cdot,\cdot)$ is Marcum's Q function, $I_0(\cdot)$ is the modified Bessel function of the first kind and order 0, and $a = \sqrt{\frac{M}{2(e^{\sigma^2}-1)} \left( 1- \sqrt{1-|r_n|^2}\right)}$, $b = \sqrt{\frac{M}{2(e^{\sigma^2}-1)} \left( 1+ \sqrt{1-|r_n|^2}\right)}$.

By now, all of the unknown terms in (\ref{MSE_final}) have been determined, and we finally have a closed-form expression of the MSE in the LSF configuration.

\subsection{The RSF Configuration}\label{theRSFconfiguration}
The $MSE(d_0|\mathbf{f})$ in (\ref{MSE_f}) is a random variable in RSF because $\mathbf{f}$ is a random vector. We choose to characterize the ranging performance of RIPS in the RSF configuration with the average of $MSE(d_0|\mathbf{f})$
\begin{equation}\label{AMSE_def}
  \overline{MSE}_{RSF} = \bar P_{o} \cdot E[(\hat{d_0}-d_0)^2|outlier],
      + (1-\bar P_{o}) \cdot \overline{CRB}_{RSF},
\end{equation}
where $\overline{MSE}_{RSF}$, $\overline{CRB}_{RSF}$, and $\bar P_{o}$ are the averages of $MSE(d_0|\mathbf{f})$, $CRB(d_0|\mathbf{f})$, and the outlier probability with respect to the random vector $\mathbf{f}$.

Rather than obtaining $\bar P_{o}$ by the traditional method in which $P_o$s with different realizations of $\mathbf{f}$ are calculated one by one, we obtain the expression of $\bar P_{o}$ immediately with the help of a concept known as the AAF, which is commonly used in random signal radars \cite{kaveh1976average}.

Averaging (\ref{v_daijiahanshu}) with respect to $\mathbf{f}$ and replacing $\varphi_i$ by (\ref{phii}) with noise free data, we get the AAF of RSF
\begin{equation}\label{AAF}
  \begin{array}{ll}
  \bar G(d)&=E\left |\sum_{i=1}^{M} \exp\left( j2\pi \frac{f_i}{c}(d_0 - d)\right)\right |\\
           &=\left| \frac{M\sin(\pi(d_0-d)\frac{Nf_{min}}{c})}{N\sin(\pi(d_0-d)\frac{f_{min}}{c})} \right|.
  \end{array}
\end{equation}
Similar to the case in LSF, we have
\begin{equation}\label{Po2}
   \bar P_{o} \approx \sum_{n=1}^{N_q} q_n
\end{equation}
and
\begin{equation}\label{out2}
  \bar P_{o} \cdot E[(\hat{d_0}-d_0)^2|outlier] \approx \sum_{n=1}^{N_q}  q_n(d'_n - d_0)^2,
\end{equation}
where $d'_n$ are positions, $N_q$ is the number of sidelobe peaks of the AAF, and $q_n = \Pr[ V(d_0)< V(d'_n)]$ is the probability that the sidelobe peak of OSF at $d'_n$ is higher than the mainlobe.

It follows from (\ref{AAF}) that $N_q =N-2$ and $d'_n = d_0 +(-1)^n \tfrac{c}{Nf_{min}}(\lceil N/2\rceil+0.5)$.

Combining (\ref{AMSE_def}), (\ref{Po2}), and (\ref{out2}), we have the closed-form expression of AMSE for RSF
\begin{equation}\label{AMSE_final}
  \overline{MSE}_{RSF} \approx \sum_{n=1}^{N_q} q_n(d'_n-d_0)^2 + \left(1 - \sum_{n=1}^{N_q}q_n\right)\overline{CRB}_{RSF},
\end{equation}
where $\overline{CRB}_{RSF}$ and $q_n$ will be determined in the following subsections.

\subsubsection{Average CRB}
Denoting $X=\mathbf{K}^{T}\mathbf{W K}$ and $g(X)=1/X$, we have
\begin{equation}\label{ACRB_def}
  \overline{CRB}_{RSF} = E\left[ \frac{M c^2 \sigma^2}{4\pi^2 f_{min}^2}\frac{1}{X}\right] = \frac{M c^2 \sigma^2}{4\pi^2 f_{min}^2} E[g(X)].
\end{equation}
The determination of $E[g(X)]$ involves the joint distribution function of the \emph{quadratic form} $\mathbf{K}^{T}\mathbf{W K}$, which is highly complex for the uniform distributed variables $k_i$ \cite{provost2013approximating}. We resort to approximations here.

Let $\eta$ be the mean and $\rho$ be the second-order moment of $X$. Expanding $g(X)$ into polynomials near $\eta$ and retaining the first three terms, we have
  $g(X) \approx g(\eta)+g'(\eta)(X-\eta)+g''(\eta)\frac{(X-\eta)^2}{2}$.
Thus $E[g(X)]$ can be approximated as
\begin{equation}\label{Eg_x}
  E[g(X)] \approx g(\eta)+\frac{g''(\eta)}{2} E[(X-\eta)^2] = \frac{\rho}{\eta ^3}.
\end{equation}

To obtain $\eta$ and $\rho$, the different orders of moment of $k_i$ should be determined first.
Considering that $N$ is a very large number because the minimum frequency interval $f_{min}$ can be as small as 1 Hz in modern transceivers \cite{rapinoja2010digital}, the $a$th-order moment of $k_i$ can be expressed as
\begin{equation}\label{ki_ju}
  \textstyle E({k_i}^a) = \frac{1}{N}\sum_{x=0}^{N-1}x^a \approx \frac{1}{N}\int_{0}^{N}x^a = \frac{N^a}{a+1}.
\end{equation}

For the two forms of quadratic terms ${k_i}^2$ and $k_ik_j(i\neq j)$ in $X$, the expectations are $E(k_i^2)$ and $E(k_i)^2$, respectively. Because the sum of coefficients of the first form is $M^2-M$ and that of the second form is $-(M^2-M)$, we have
\begin{equation}\label{eta1}
\begin{array}{ll}
  \eta &= (M^2-M) E({k_i}^2) - (M^2-M)E(k_i)^2 \\
       &= \frac{M(M-1)N^2}{12}.
\end{array}
\end{equation}

Similarly, $\rho$ is the sum of expectations of various quartic terms. These expectations have five forms: $E(k_i^4)$, $E(k_i)E(k_j^3)$, $E(k_i^2)^2$, $E(k_i^2) E(k_j)^2$ and $E(k_i)^4$ where $i\neq j$. The sums of coefficients of these five forms are $\beta_1=M(M-1)^2$, $\beta_2=-4M(M-1)^2$, $\beta_3=M(M-1)[(M-1)^2+2]$, $\beta_4=-2M(M-1)(M-2)(M-3)$, and $\beta_5=M(M-1)(M-2)(M-3)$. Therefore,
\begin{equation}\label{eta2}
\begin{array}{ll}
  \rho &= \beta_1 E(k_i^4)+ \beta_2 E(k_i)E(k_j^3) + \beta_3 E(k_i^2)^2\\
   &+\beta_4 E(k_i^2) E(k_i)^2+ \beta_5 E(k_i)^4\\
   &= \frac{M(M-1)(5M^2-M+6)N^4}{720}.
\end{array}
\end{equation}

Plugging (\ref{eta1}) and (\ref{eta2}) back into (\ref{ACRB_def}) and (\ref{Eg_x}), we have
\begin{equation}\label{ACRB_final}
  \overline{CRB}_{RSF} \approx \frac{3 c^2 \sigma^2}{5 \pi^2 B^2}\frac{(5M^2 -M +6)}{M(M-1)^2}.
\end{equation}

\subsubsection{The Determination of $q_n$}
For RSF, the means and second-order moments of $y_0$ and $y_n$ are
\begin{equation}\label{moments2}
  \begin{array}{ll}
    E[y_0] = M e^{-\sigma^2/2}\\
    E[y_n] = M r_n e^{-\sigma^2/2}\\
    var[y_0] = 0.5M (1-e^{-\sigma^2})\\
    var[y_n] = 0.5M (1-|r_n|^2 e^{-\sigma^2})\\
    cov[y_0,y_n] = 0.5M ({r_n})^* (1-e^{-\sigma^2}),
  \end{array}
\end{equation}
where $r_n = E\left[ \exp \left( j2\pi \frac{f_i}{c}(d_0-d'_n)\right)\right]$.

Substituting (\ref{moments2}) into B-21 of \cite{proakis}, we have
\begin{equation}\label{qn}
  q_n = Q_1(c,d) - v \cdot I_0(cd) \exp\left[-\tfrac{1}{2}(c^2+d^2)\right],
\end{equation}
where $c = \sqrt{\frac{M}{A_n+S} \left( 2e^{\sigma^2} - A_n - \sqrt{{A_n}^2+A_nS}\right)}$, $d = \sqrt{\frac{M}{A_n+S} \left( 2e^{\sigma^2} - A_n + \sqrt{{A_n}^2+A_nS}\right)}$, $v = \frac{\sqrt{1+S/A_n}-1}{2\sqrt{1+S/A_n}}$, of which $A_n= 1 -|r_n|^2$ and $S = 4(e^{2\sigma^2}-e^{\sigma^2})$.

Hence, we have obtained in (\ref{AMSE_final}) a closed-form expression of AMSE in the RSF configuration.

\section{Simulation Results}
In this section, the accuracy of the approximations derived in the previous section is verified through Monte Carlo simulations. For a fair comparison, we assume that LSF and RSF employ the same frequency band with a bandwidth of $B= 0$ MHz and the same minimum frequency interval $f_{min}=1$ kHz. The number of Monte Carlo trials was $10^5$ for each SNR. In each trial, we use $M$ measurement frequencies for LSF and randomly choose $M$ frequencies for RSF.

Fig.1 shows the MSE of the ML estimator as a function of SNR when $M$ is 31. The MSE prediction is hard limited to never exceed the variance $d_{max}^2/12$ of an estimate assumed to be uniformly distributed over the search space.

\begin{figure}[h]
       \centering
       \includegraphics[width=5in]{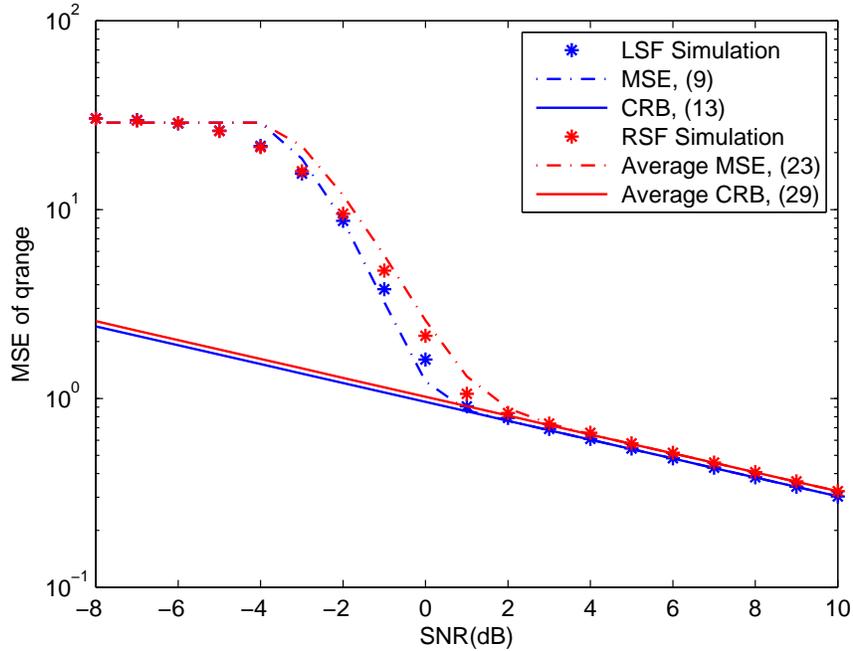}
       \caption{Monte Carlo simulations of the ML qrange estimator in the LSF configuration and RSF configuration with $B=30$ MHz and $M=31$. The number of Monte Carlo trials was $10^5$.
       Blue: the MSE approximation of LSF is compared with simulation and CRB.
       Red: the AMSE approximation of RSF is compared with simulation and Average CRB.
       }
\end{figure}

We determined that both the derived MSE approximation in (\ref{MSE_final}) for LSF and the AMSE approximation in (\ref{AMSE_final}) for RSF are accurate. Moreover, the AMSE in RSF is only slightly larger than the MSE in LSF in the entire SNR region. For military applications, RSF is superior to LSF because the tiny loss of accuracy in RSF compared with LSF is well compensated by its anti-jamming capabilities.

\section{Conclusions}
This letter provides closed-form expressions of the MSE of the ML qrange estimator for the RIPS in the LSF configuration and the AMSE in the RSF configuration. The simulation results agree well with the theoretical analysis. We conclude that RSF increases the anti-jamming capability of RIPS at a very small cost of slightly decreased ranging accuracy.


%

\ifCLASSOPTIONcaptionsoff
  \newpage
\fi





\begin{thebibliography}{}

\bibitem{maroti2005radio}
M. Maroti, P. Volgyesi, S. Dora, B. Kusy, A. Nadas, A. Ledeczi, G. Balogh, and K. Molnar, ``Radio interferometric
geolocation," in \textit{Proc. the 3rd international conference on Embedded networked sensor systems.} ACM, 2005, pp. 1--12.

\bibitem{wang2013design}
Y. Wang, M. Shinotsuka, X. Ma, and M. Tao, ``Design an asynchronous radio interferometric positioning system using
dual-tone signaling," in \textit{Proc. Wireless Communications and Networking Conference.} IEEE, 2013, pp. 2294--2298.

\bibitem{li2013distance}
W. Li, X. Wang, X. Wang, and B. Moran, ``Distance estimation using wrapped phase measurements in noise," \textit{IEEE Trans.
Signal Process.}, vol. 61, no. 7, pp. 1676--1688, 2013.

\bibitem{kusy2007intrack}
B. Kusy, G. Balogh, J. Sallai, A. Ledeczi, and M. Maroti, ``intrack: High precision tracking of mobile sensor nodes," in
\textit{Wireless Sensor Networks}. Springer, 2007, pp. 51--66.

\bibitem{liu2011ground}
P. Liu, W. D. Qi, E. Yuan, Y. S. Zhu, and H. Wang, ``Ground displacement measurement by radio interferometric ranging
for landslide early warning," in \textit{Proc. Instrumentation and Measurement Technology Conference.} IEEE, 2011, pp. 1--6.

\bibitem{weili}
L. Wei, W. Qi, P. Liu, E. Yuan, Y. Zhu, and X. Ji, ``Method for selecting measurement frequencies based on dual pseudorandom
code in radio interferometric positioning system," China Patent CN102 221 695, June 12, 2013.

\bibitem{athley2005threshold}
F. Athley, ``Threshold region performance of maximum likelihood direction of arrival estimators," \textit{IEEE Trans. Signal
Process.}, vol. 53, no. 4, pp. 1359--1373, 2005.

\bibitem{richmond2006mean}
C. D. Richmond, ``Mean-squared error and threshold snr prediction of maximum-likelihood signal parameter estimation
with estimated colored noise covariances," \textit{IEEE Trans. Inf. Theory}, vol. 52, no. 5, pp. 2146--2164, 2006.

\bibitem{MIMOradar07}
N. H. Lehmann, E. Fishler, A. M. Haimovich, R. S. Blum, D. Chizhik, L. J. Cimini, and R. A. Valenzuela, ``Evaluation
of transmit diversity in mimo-radar direction finding," \textit{IEEE Trans. Signal Process.}, vol. 55, no. 5, pp. 2215--2225, May
2007.

\bibitem{kaveh1976average} 
M. Kaveh and G. R. Cooper, ``Average ambiguity function for a randomly staggered pulse sequence," \textit{IEEE Trans. Aerosp.
Electron. Syst.}, vol. 12, no. 3, pp. 410--413, 1976

\bibitem{tretter1985estimating}
S. Tretter, ``Estimating the frequency of a noisy sinusoid by linear regression," \textit{IEEE Trans. Inf. Theory}, vol. 31, no. 6, pp.
832--835, 1985.

\bibitem{awoseyila2008improved}
A. B. Awoseyila, C. Kasparis, and B. G. Evans, ``Improved single frequency estimation with wide acquisition range,"
\textit{Electron. Lett.}, vol. 44, no. 3, pp. 245--247, 2008.

\bibitem{rife1974}
D. C. Rife and R. Boorstyn, ``Single tone parameter estimation from discrete-time observations," \textit{IEEE Trans. Inf. Theory},vol. 20, no. 5, pp. 591--598, 1974.

\bibitem{proakis}
J. G. Proakis, \textit{Digital Communications}, 4th, Ed. McGraw-Hill, New York, 2001.

\bibitem{papoulis}
A. Papoulis and S. U. Pillai, \textit{Probability, Random Variables, and Stochastic Processes}, 4th, Ed. New York: McGraw-Hill,
2002.

\bibitem{provost2013approximating}
S. B. Provost and A. A. Mohsenipour, ``On approximating the distribution of quadratic forms in uniform and beta order
statistics," \textit{Metron}, vol. 71, no. 2, pp. 123--138, 2013.

\bibitem{rapinoja2010digital}
T. Rapinoja, K. Stadius, L. Xu, S. Lindfors, R. Kaunisto, A. Parssinen, and J. Ryynanen, ``A digital frequency synthesizer
for cognitive radio spectrum sensing applications," \textit{IEEE Trans. Microw. Theory Tech.}, vol. 58, no. 5, pp. 1339--1348, 2010.

\end{thebibliography}

\end{document}